\documentclass{INTERSPEECH2023}

\interspeechcameraready

\title{Efficient Encoder-Decoder and Dual-Path Conformer for Comprehensive Feature Learning in Speech Enhancement}
\name{Junyu Wang}
\address{
  School of Electronic Information, Sichuan University, China}
\email{junyu\_wang@stu.scu.edu.cn}

\begin{document}

\maketitle
 
\begin{abstract}
Current speech enhancement (SE) research has largely neglected channel attention and spatial attention, and encoder-decoder architecture-based networks have not adequately considered how to provide efficient inputs to the intermediate enhancement layer. To address these issues, this paper proposes a time-frequency (T-F) domain SE network (DPCFCS-Net) that incorporates improved densely connected blocks, dual-path modules, convolution-augmented transformers (conformers), channel attention, and spatial attention. Compared with previous models, our proposed model has a more efficient encoder-decoder and can learn comprehensive features. Experimental results on the VCTK+DEMAND dataset demonstrate that our method outperforms existing techniques in SE performance. Furthermore, the improved densely connected block and two dimensions attention module developed in this work are highly adaptable and easily integrated into existing networks. 
\end{abstract}

\section{Introduction}

Speech enhancement (SE) is crucial in many speech processing systems, including but not limited to speaker verification, hearing aids, and automatic speech recognition. Single-channel SE remains a challenging issue due to its reliance on a single speech signal for separation and enhancement. Traditional signal-processing-based SE methods struggle to effectively separate and eliminate interference in actual speech signals, such as noise and reverberation, resulting in poor enhancement results. In light of the recent success of deep learning in diverse domains \cite{interspeechconvtasnet, interspeech24}, SE employing deep neural networks (DNNs) has emerged as the prevailing approach. Based on the processing of the input signal, the technique can typically be divided into time-frequency (T-F) domain methods and time-domain methods. This paper mainly focuses on the T-F domain single-channel SE.

Specifically, the time-domain approach focuses on estimating clean speech samples directly from the original data of noisy speech samples. Instead of fixed T-F domain transformations, this approach employs a trainable encoder and decoder to avoid computation related to converting between the time and frequency domains \cite{interspeech3, interspeech4, interspeech6}. Nevertheless, the simplicity of time-domain features may lead to a lack of interpretability in the generated features, which limits the performance of time-domain methods to some extent.

Most methods in the T-F domain utilize the STFT to extract input features and training targets. Compared to the encoder and decoder that work on a short signal window in the time-domain approach, T-F domain models typically use a larger window size and stride size and have stronger robustness. However, the current SE methods primarily focus on enhancing the spectrum magnitude and utilizing the phase of the noise mixture for signal reconstruction \cite{interspeech8, interspeech10, interspeech35}, limiting the upper limit of enhancement performance. Some recent studies have considered complex spectra features that maximize the preservation of phase information, achieving better performance \cite{interspeech11, interspeech12}.

Dual-path network (DPN) is a widely utilized mechanism in various areas of speech signal processing, for example, in speech separation \cite{interspeech14} and speech recognition \cite{interspeech15}. It was initially introduced in \cite{interspeech14} and has demonstrated exceptional performance by alternately learning local and global features of sequences. In recent years, numerous studies \cite{interspeech17, interspeech18, interspeech19} have integrated the DPN with the transformer architecture, enabling direct interaction between input elements and facilitating the modelling of longer speech sequences, thereby further enhancing the performance of the DPN. However, these studies primarily focus on local and global information of input signals without considering the integration of channel and spatial information with the existing DPN. Furthermore, we observe that most of these methods fail to consider how to provide adequate inputs to the enhancement layer in the network. \cite{interspeech17} attempted to extend the encoder and decoder functions by introducing densely connected blocks \cite{interspeech22}, but the resulting performance improvement is modest.

In light of the issues above, we present a T-F domain SE network (DPCFCS-Net) based on deep connection blocks (DCBs), dual-path modules, convolution-augmented transformers (conformers) \cite{interspeech20, interspeech21}, channel attention, and spatial attention. The network is composed of an encoder, an enhancement layer, and a decoder. The encoder and decoder are based on our proposed DCB and two dimensions attention module, aimed at providing efficient input feature space to the enhancement layer. The enhancement layer comprises dual-path modules and conformers. Using the input enhanced with the representation of specific regions through the two dimensions attention module, the dual-path conformer (DPCF) can better alternate learning of sub-band and full-band information. In general, our contributions are as follows:

\begin{itemize}

\item We design a DCB to extract rich and accurate features, which outperforms the densely connected block in performance under lower computational complexity.
\item We introduce a module combining spatial and channel attention to improve the feature learning ability of the network, which can easily integrate into existing networks.
\item We propose a T-F domain SE network by integrating dual-path modules, conformers, DCBs, and channel and spatial attention to enable the model to learn comprehensive features.
\item Extensive experiments on the VCTK+DEMAND dataset \cite{interspeech23} demonstrate that the SE performance of our proposed model outperforms all current models.

\end{itemize}

\section{Deep connection block and two dimensions attention module}

\subsection{Deep connection block}

The existing densely connected block is an improvement derived from ResNets \cite{interspeech24}, which enhances the flow of information between layers by introducing direct connections from all previous to subsequent layers, allowing the \(n^{th}\) layer to receive information from all preceding layers in total \(n-1\). However, with the increasing number of network layers, the amount of input information received by later layers increases greatly, significantly increasing the training burden.

To address this issue, we present a light-weighted connection block (LWCB) and a DCB to replace the densely connected block. By utilizing direct connections from the initial layer to all successive layers instead of from all previous layers to subsequent layers, the proposed LWCB not only improves the flow of information within the network but also avoids an exponential increase in computational complexity with increasing network layers, as shown in Figure 1(a). Inspired by the U-Net structure \cite{interspeech25} and multi-scale fusion \cite{interspeech26}, we appropriately combine \cite{interspeech25, interspeech26} with the LWCB to generate the DCB, as shown in Figure 1(b). The information initially flows from bottom to top through the various stages of the LWCB, then fuses with the adjacent blocks above to facilitate the mutual learning of adjacent stages, and finally merges at the bottom layer with skip connections. Compared to the single transmission of information from bottom to top in the densely connected block, the information transmission in the DCB starts from the bottom, reaches the top, returns to the bottom, and merges information from different stages, which is more conducive to learning comprehensive features.

\subsection{Two dimensions attention module}

The attention mechanism has emerged as a crucial component of modern deep learning models, enabling selective focus on specific portions of the input data instead of treating the entire input uniformly. Appropriate utilization of attention mechanisms can result in reduced computational complexity and improved network performance. Inspired by the successful utilization of channel and spatial attention in computer vision \cite{interspeech27, interspeech28}, we propose a two dimensions attention module for SE, as depicted in Figure 2.

 We first aggregate speech information by applying 2-D max pooling and 2-D average pooling. Subsequently, a 1-D convolution operation is employed to generate the mapping of channel attention, which can be represented as:
\begin{align}
  AT_c(E) &= \delta(C_{1d}(MP_{2d}(E))+C_{1d}(AP_{2d}(E))) 
\end{align}
where \(AT_c()\) is the mapping of channel attention, \(\delta\) denotes the sigmoid activation function, \(C_{1d}()\) represents a 1-D convolution, and \(MP_{2d}()\) and \(AP_{2d}()\) denote 2-D max pooling and 2-D average pooling, respectively.

The output \(E'\) of channel attention first aggregates speech channel information through 1-D max pooling and 1-D average pooling, followed by using a 2-D convolution layer to generate the mapping of spatial attention, which can be represented as:
\begin{align}
  AT_s(E') &= \delta(C_{2d}([MP_{1d}(E'); AP_{1d}(E')])) 
\end{align}
where \(AT_s()\) is the mapping of spatial attention, \(\delta\) denotes the sigmoid activation function, \(C_{2d}()\) represents a 2-D convolution, and \(MP_{1d}()\) and \(AP_{1d}()\) denote 1-D max pooling and 1-D average pooling, respectively.

\begin{figure}[t]
  \centering
  \includegraphics[width=\linewidth]{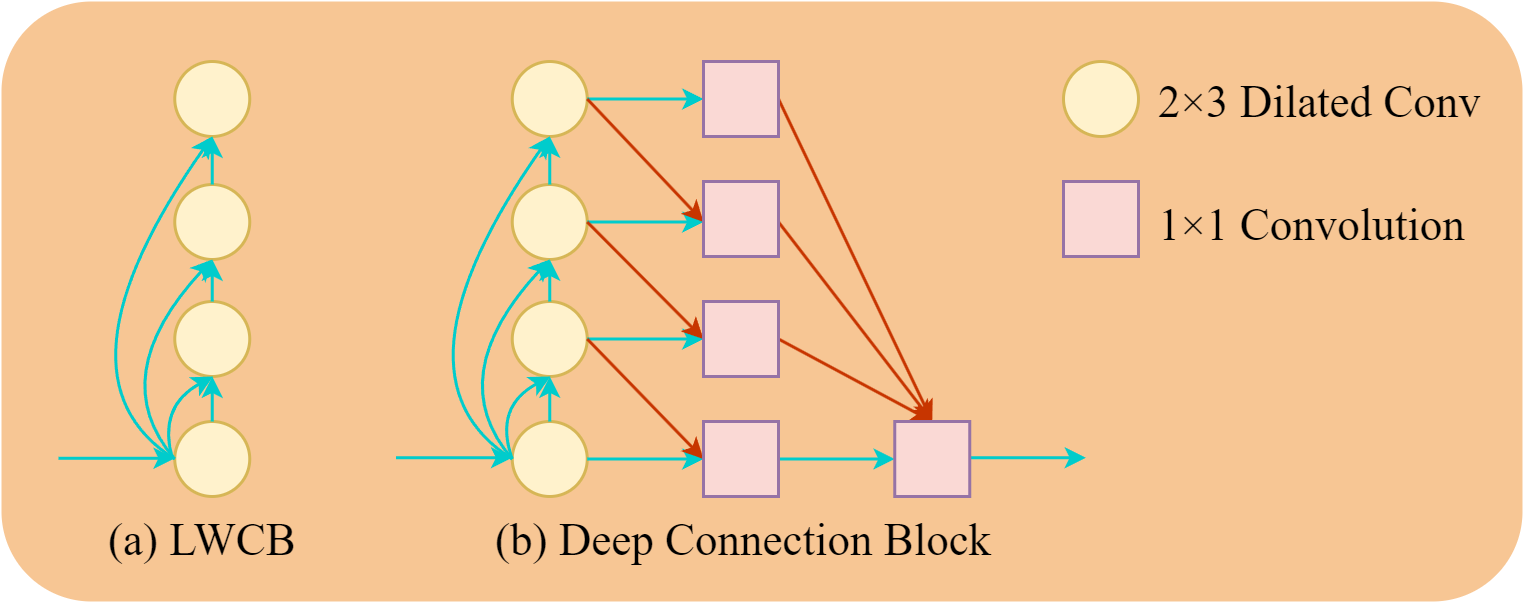}
  \caption{Deep connection block and light-weighted connection block.}
  \label{figure1}
\end{figure}

\begin{figure}[t]
  \centering
  \includegraphics[width=\linewidth]{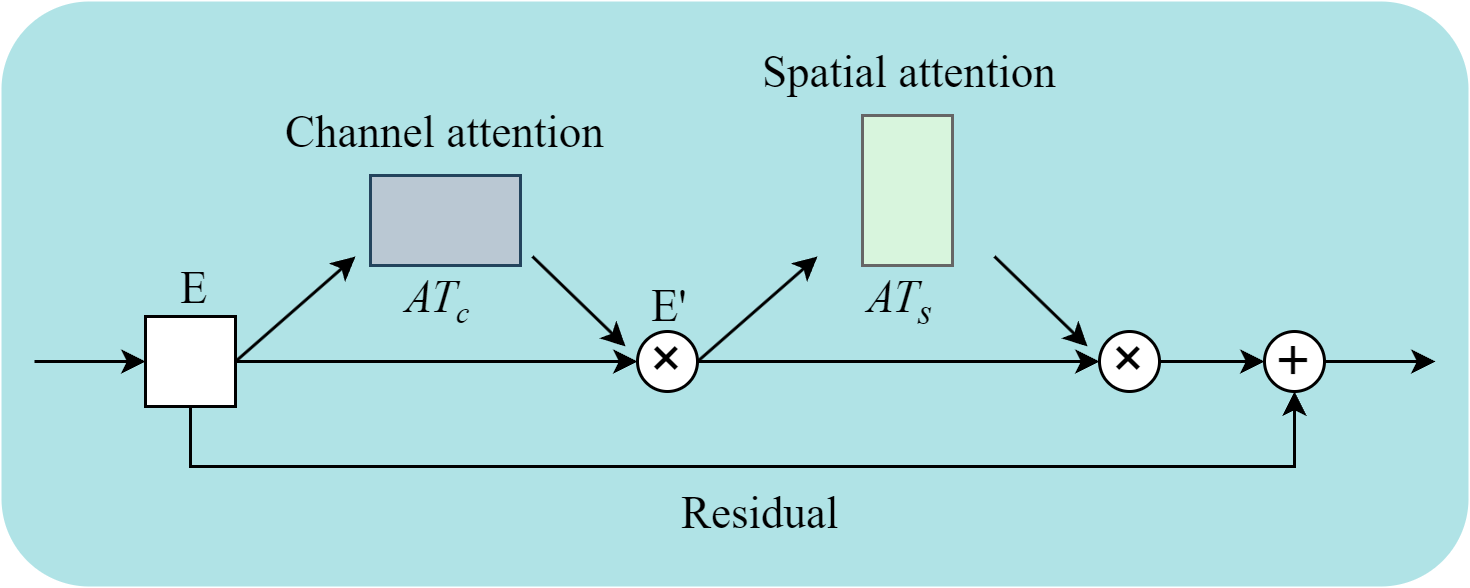}
  \caption{Two dimensions attention module.}
  \label{figure1}
\end{figure}

\begin{figure*}[t]
  \centering
  \includegraphics[width=0.7\linewidth]{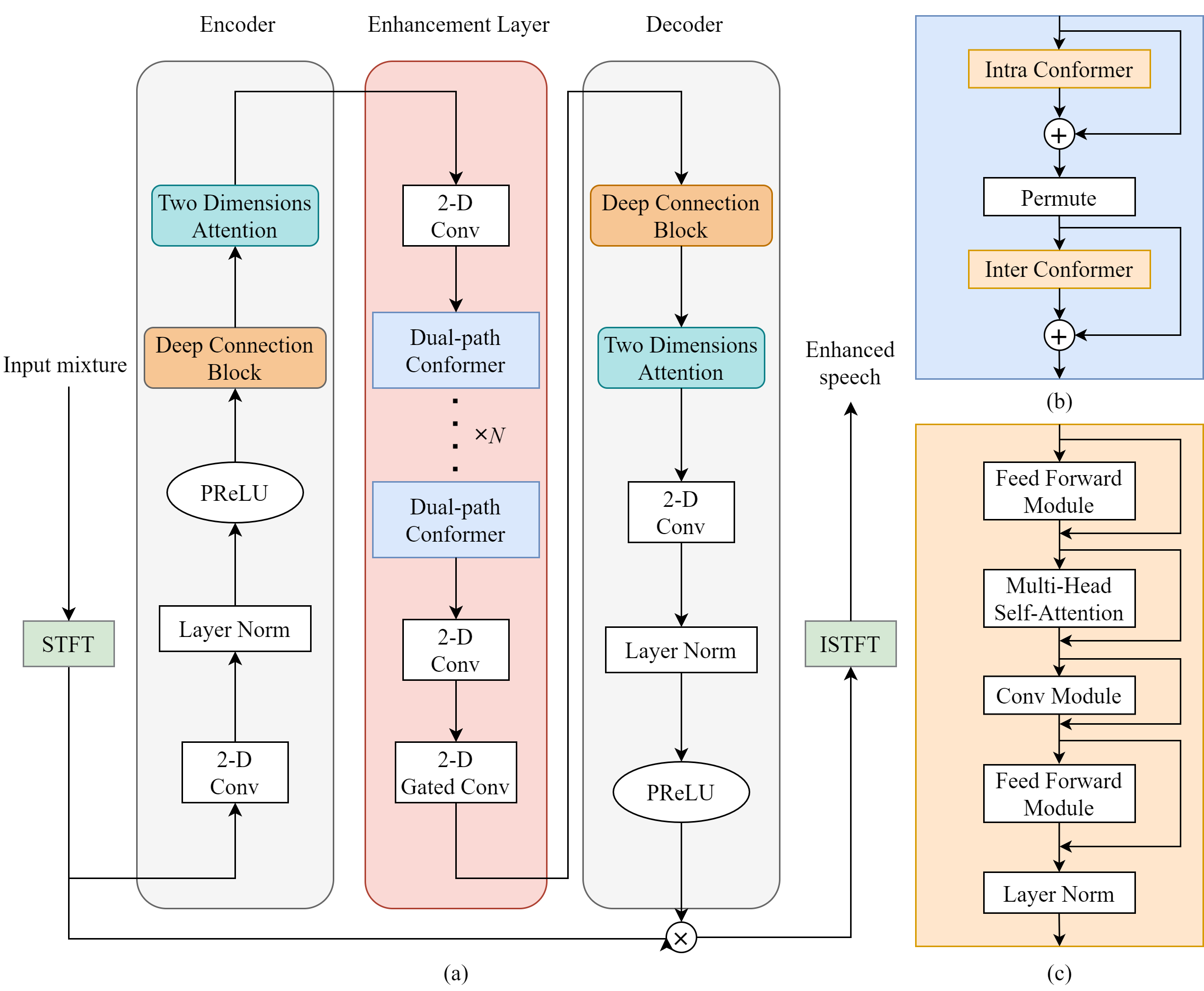}
  \caption{Structure diagram of DPCFCS-Net. (a) Overall illustration. (b) Dual-path conformer module. (c) Conformer block.}
  \label{figure3}
\end{figure*}

\section{DPCFCS-Net}

Our proposed model, as illustrated in Figure 3, comprises three phases: encoder, decoder, and enhancement layer. During the encoder phase, the mixed waveform segments are transformed into corresponding features in the intermediate feature space. The enhancement layer then receives the features and constructs masks for both the clean speech and noise sources. Finally, in the decoder phase, the masked features are transformed back into the source waveform \cite{interspeech18}.

\subsection{Encoder}

The encoder comprises a DCB, a two dimensions attention module, and a 1\(\times\)1 convolutional layer. Initially, the input complex spectrogram generated by STFT undergoes convolutional layer processing to boost the channel count from 2 to 128. Then it passed through the DCB for adequate feature extraction. Finally, the two dimensions attention module provides a more efficient representation of the input features to the following enhancement layer. The inputs and outputs of the encoder are denoted as \(X \in \mathbb{R}^{2 \times T \times F}\) and \(U \in \mathbb{R}^{128 \times T \times F}\), respectively. 

\subsection{Enhancement layer}

The enhancement layer comprises two 1\(\times\)1 convolutional layers, \(N\) DPCFs, and a 1\(\times\)1 gated convolutional layer. Both convolutional layers are followed by layer normalization and SMU activation \cite{interspeechsmu}. The output of the encoder through the first convolutional layer is halved in channel dimension to obtain \(P \in \mathbb{R}^{64 \times T \times F}\), which is used as input for the DPCF. The DPCF comprises an intra-conformer and an inter-conformer, which respectively model sub-band and full-band information \cite{interspeech29}. The conformer block, as depicted in Figure 3(c), consists of feedforward (FFN) modules, convolution module, multi-head self-attention (MHSA) module, and layer normalization \cite{interspeech21}. 

The intra-conformer block first processes the input \(P\) to model the sub-band features in the block length dimension \(T\). Then, the inter-conformer block aggregates all sub-bands information output from the intra-conformer to learn full-band information, which acts on the dimension of the number of blocks \(F\). Subsequently, the last 1\(\times\)1 convolutional layer doubles the channel dimension of the output of the DPCF to obtain \(D \in \mathbb{R}^{128 \times T \times F}\). Finally, the 2-D gated convolutional layer smooths the enhancement layer output value.

\subsection{Decoder}

The decoder comprises the same DCB and two dimensions attention module as the encoder, as well as a 1\(\times\)1 convolution layer. After passing through the enhancement layer, the output is fed into the decoder to obtain the decoded mask features. To restore the channel dimension of the mask, a 1\(\times\)1 convolution layer is employed. The enhanced complex spectrogram of the speech is computed by element-wise multiplication of the STFT output and the mask, and the final speech waveform is obtained through the application of ISTFT.

\subsection{Loss function}

The loss function comprises both speech \(x\) and noise \(n\) losses \cite{interspeech30}, defined as follows:
\begin{align}
  loss_{total}(x,n) &= \alpha loss(x,\hat{x})+(1-\alpha)loss(n,\hat{n})
\end{align}
where \(\hat{x}\) and \(\hat{n}\) represent the estimated speech and noise, respectively, and \(\alpha\) is defined as \(\alpha = \Vert x\Vert^{2}/(\Vert x\Vert^{2}+\Vert n\Vert^{2})\).

We also apply both time-domain loss (MSE) and T-F domain loss (L1) \cite{interspeech29} to optimize the model, as described by the following formula:
\begin{equation}
\begin{aligned}
loss(x,\hat{x}) &= \beta loss(x,\hat{x})_{T}+(1-\beta) loss(x,\hat{x})_{F} \\
l&oss(x,\hat{x})_{T} = \frac{1}{M}\sum\limits_{v=0}^{M-1}(x_{v}-\hat{x}_{v})^2 \\
loss(x,\hat{x})_{F} &= \frac{1}{TF}\sum\limits_{t=0}^{T-1}\sum\limits_{f=0}^{F-1}\vert (\vert X_{r}(t,f)\vert-\vert \hat{X}_{r}(t,f)\vert) \\
&+(\vert X_{i}(t,f)\vert-\vert \hat{X}_{i}(t,f)\vert)\vert
\end{aligned}
\end{equation}
where hyperparameter \(\beta\) is set to 0.4. \(T\) and \(F\) represent the number of frames and frequency bins. \(X\) and \(\hat{X}\) represent the spectrogram of the clean speech and enhanced speech. \(r\) and \(i\) refer to the real and imaginary components. \(M\) denotes the number of samples in the waveform.

\begin{table*}[ht]
\begin{center}
  \caption{Performance comparison with other baseline models on the VCTK dataset. "-" represents the data not provided in the paper}
  \setlength{\tabcolsep}{4mm}
  \label{tab1}
  \centering
  \begin{tabular}{cccccccc}
    \hline
    \textbf{Model}  & \textbf{Domain}   & \textbf{PESQ}  & \textbf{CSIG}   & \textbf{CBAK}   & \textbf{COVL}  & \textbf{STOI}   & \textbf{Params.(M)}  \\
    \hline
    Noisy  & - & 1.97 & 3.35 & 2.44 & 2.63 & 0.91 & -                 \\
    \hline
    SEGAN \cite{interspeech3}  & T  & 2.16 & 3.48 & 2.94 & 2.80 & 0.92 & 97.47                                \\
    MetricGAN \cite{interspeech35}  & T  & 2.86 & 3.99 & 3.18 & 3.42 & - & -                          \\
    PHASEN \cite{interspeech36} & F & 2.99 & 4.21 & 3.55 & 3.62 & - & 8.76                               \\
    TSTNN \cite{interspeech17} & T & 2.96 & 4.10 & 3.77 & 3.52 & 0.95 & 0.92                               \\
    MSSA-TCN \cite{interspeech37}  & F  & 3.02 & 4.29 & 3.50 & 3.67 & 0.94 & 9.91                          \\
    DEMUCS \cite{interspeech6} & T & 3.07 & 4.31 & 3.40 & 3.63 & 0.95 & 33.5                    \\
    SE-Conformer \cite{interspeech21} & T & 3.13 & 4.45 & 3.55 & 3.82 & 0.95 & -                 \\
    DPT-FSNet \cite{interspeech29} & F & 3.33 & 4.58 & 3.72 & 4.00 & 0.96 & 0.91                 \\
    DB-AIAT \cite{interspeech12} & F & 3.31 & 4.61 & 3.75 & 3.96 & 0.96 & 2.81                 \\
    \hline
    \textbf{DPCFCS-Net}  & F & \textbf{3.42} & \textbf{4.71} & \textbf{3.88} & \textbf{4.15} & \textbf{0.96} & 2.86                 \\
    \hline
  \end{tabular}
\end{center}
\end{table*}

\section{Experiments}

\subsection{Dataset}

To assess the efficacy of the presented model for SE under different noise conditions, we use the widely used VCTK+DEMAND dataset \cite{interspeech23}. The dataset is a mixture of the DEMAND dataset \cite{interspeechDemand} and the VoickBank corpus \cite{interspeechvoice}. The clean speech data comprises 12396 utterances from 30 speakers (15 male and 15 female) in the VoickBank corpus, with 28 speakers utilized for training and 2 speakers utilized for testing. The training set for the noise set comprises 10 types of noise, including 2 artificially generated noise types and 8 from the DEMAND dataset, with SNRs of 15, 10, 5, and 0 dB. The test set includes 5 different types of noise, with SNRs of 17.5, 12.5, 7.5, and 2.5 dB. All speech samples are downsampled to 16kHz.

\subsection{Experimental setup}

The window length and hop size are 25 ms and 6.25 ms, with the FFT length of 512. In the DCB, layer normalization and SMU activation are applied after all convolution operations, and there are four dilated convolutions with the dilation rate increases by a multiple of two from low to high. The number of DPCFs is 4 (\(N=4\)), and each conformer block has four attention heads. During training, we randomly slice speech samples into 4-second segments. The proposed model undergoes 150 epochs of training, optimized by AdamW \cite{interspeech31}, with an initial learning rate of \(5e^{-4}\) that decays by a factor of 0.95 every four epochs.

\subsection{Evaluation metrics}

We assess the effectiveness of the presented model using several subjective and objective metrics. STOI is used to measure speech intelligibility \cite{interspeech32}, with a rating of 0 to 1. PESQ is used to evaluate speech quality \cite{interspeech33}, which scores from -0.5 to 4.5. For the mean opinion score (MOS) rating scales, the CSIG score is the MOS predictor for signal distortion, the CBAK score is the MOS predictor for background noise interference, and the COVL score is the MOS predictor for overall speech quality, with all MOS scores range from 1 to 5.

\section{Results}

\subsection{Performance comparison on VCTK}

Table~\ref{tab1} presents the performance comparison of our proposed model DPCFCS-Net with previous models. The results indicate that our approach outperforms all other models in all evaluation metrics with fewer model parameters.

\subsection{Ablation study}

We conduct three experiments for comparison, among which DPCFN represents the model without the DCB and two dimensions attention module. The results in Table~\ref{tab2} show that correctly applying the DCB and two dimensions attention module can improve the performance of the network.

\begin{table}[h]
  \caption{Comparison of ablation experiment results}
  \label{tab2}
  \centering
  \tabcolsep=0.095cm
  \begin{tabular*}{\hsize}{@{}@{\extracolsep{\fill}}cccccc@{}}
    \hline
    \textbf{Model} & \textbf{PESQ} & \textbf{CSIG} & \textbf{CBAK} & \textbf{COVL}  & \textbf{STOI}  \\
    \hline
    DPCFN  & 3.25 & 4.53 & 3.67 & 3.95 & 0.95                              \\
    DPCFN+DCB  & 3.37 & 4.66 & 3.83 & 4.09 & 0.96                        \\
    DPCFCS-Net & \textbf{3.42} & \textbf{4.71} & \textbf{3.88} & \textbf{4.15}  & \textbf{0.96}               \\
    \hline
  \end{tabular*}
\end{table}

\subsection{The influence of deep connection block and two dimensions attention module}

To further validate the advantage of the proposed DCB and two dimensions attention module in feature learning, we choose the DDAEC model proposed in \cite{interspeech4} as a baseline, which consists of multiple densely connected blocks as the backbone network. Based on DDAEC, we design a comparison model that only replaces the densely connected blocks used in DDAEC with the proposed DCBs and two dimensions attention modules. Except for using the VCTK dataset, all other model parameters and experimental settings are kept the same as in \cite{interspeech4}. Table~\ref{tab3} shows that the comparison model outperforms DDAEC in all evaluation metrics while having a smaller model size. This further demonstrates the effectiveness of the DCB and two dimensions attention module.

\begin{table}[h]
  \caption{Performance of comparison model and DDAEC}
  \label{tab3}
  \centering
  \tabcolsep=0.10cm
  \begin{tabular*}{\hsize}{@{}@{\extracolsep{\fill}}ccccccc@{}}
    \hline
    \textbf{Model} & \textbf{PESQ} & \textbf{CSIG} & \textbf{CBAK} & \textbf{COVL} & \textbf{STOI} & \textbf{Par.(M)}  \\
    \hline
    DDAEC  & 2.23 & 3.80 & 2.57 & 3.01 & 0.90 & 4.82                               \\
    Comparison  & \textbf{2.32} & \textbf{3.91} & \textbf{2.64} & \textbf{3.10} & \textbf{0.92} & \textbf{3.55}                       \\
    \hline
  \end{tabular*}
\end{table}

\section{Conclusions}

This paper presents a novel DPCFCS-Net for speech enhancement based on dual-path conformers, deep connection blocks, and two dimensions attention modules. Experiments on the VCTK dataset demonstrate that the presented model outperforms other baselines on various performance evaluation metrics. Additionally, we show that our deep connection block and two dimensions attention module have the advantages of fewer parameters and better performance compared to the widely used densely connected block in speech enhancement. In the future, our research will consider expanding to other tasks, such as multi-channel and dereverberation.

\bibliographystyle{IEEEtran}
\bibliography{mybib}

\begin{thebibliography}{10}
\providecommand{\url}[1]{#1}
\csname url@samestyle\endcsname
\providecommand{\newblock}{\relax}
\providecommand{\bibinfo}[2]{#2}
\providecommand{\BIBentrySTDinterwordspacing}{\spaceskip=0pt\relax}
\providecommand{\BIBentryALTinterwordstretchfactor}{4}
\providecommand{\BIBentryALTinterwordspacing}{\spaceskip=\fontdimen2\font plus
\BIBentryALTinterwordstretchfactor\fontdimen3\font minus
  \fontdimen4\font\relax}
\providecommand{\BIBforeignlanguage}[2]{{%
\expandafter\ifx\csname l@#1\endcsname\relax
\typeout{** WARNING: IEEEtran.bst: No hyphenation pattern has been}%
\typeout{** loaded for the language `#1'. Using the pattern for}%
\typeout{** the default language instead.}%
\else
\language=\csname l@#1\endcsname
\fi
#2}}
\providecommand{\BIBdecl}{\relax}
\BIBdecl

\bibitem{interspeechconvtasnet}
Y.~Luo and N.~Mesgarani, ``Conv-tasnet: Surpassing ideal time-frequency
  magnitude masking for speech separation,'' \emph{IEEE/ACM Transactions on
  Audio, Speech, and Language Processing}, vol.~27, no.~8, pp. 1256--1266,
  2019.

\bibitem{interspeech24}
K.~He, X.~Zhang, S.~Ren, and J.~Sun, ``Deep residual learning for image
  recognition,'' in \emph{IEEE Conference on Computer Vision and Pattern
  Recognition (CVPR)}, {Las Vegas, NV, USA}, 2017, pp. 770--778.

\bibitem{interspeech3}
S.~Pascual, A.~Bonafonte, and J.~Serrà, ``Segan: Speech enhancement generative
  adversarial network,'' in \emph{Proc. {INTERSPEECH} 2017 --
  18\textsuperscript{th} Annual Conference of the International Speech
  Communication Association}, {Stockholm, Sweden}, 2017, pp. 3642--3646.

\bibitem{interspeech4}
A.~Pandey and D.~L. Wang, ``Densely connected neural network with dilated
  convolutions for real-time speech enhancement in the time domain,'' in
  \emph{IEEE International Conference on Acoustics, Speech and Signal
  Processing (ICASSP)}, {Barcelona, Spain}, 2020, pp. 6629--6633.

\bibitem{interspeech6}
A.~Defossez, G.~Synnaeve, and Y.~. Adi, ``Real time speech enhancement in the
  waveform domain,'' in \emph{Proc. {INTERSPEECH} 2020 --
  21\textsuperscript{st} Annual Conference of the International Speech
  Communication Association}, {Shanghai, China}, 2020, pp. 3291--3295.

\bibitem{interspeech8}
Y.~Xu, J.~Du, L.-R. Dai, and C.-H. Lee, ``A regression approach to speech
  enhancement based on deep neural networks,'' \emph{IEEE/ACM Transactions on
  Audio, Speech, and Language Processing}, vol.~23, no.~1, pp. 7--19, 2015.

\bibitem{interspeech10}
K.~Tan, J.~Chen, and D.~L. Wang, ``Gated residual networks with dilated
  convolutions for supervised speech separation,'' in \emph{IEEE International
  Conference on Acoustics, Speech and Signal Processing (ICASSP)}, {Calgary,
  Canada}, 2018, pp. 21--25.

\bibitem{interspeech35}
S.-W. Fu, C.-F. Liao, Y.~Tsao, and S.-D. Lin, ``Metricgan: Generative
  adversarial networks based black-box metric scores optimization for speech
  enhancement,'' in \emph{International Conference on Machine Learning (ICML)},
  2019, pp. 2031--2041.

\bibitem{interspeech11}
A.~Li \emph{et~al.}, ``Two heads are better than one: A two-stage complex
  spectral mapping approach for monaural speech enhancement,'' \emph{IEEE/ACM
  Transactions on Audio, Speech, and Language Processing}, vol.~29, pp.
  1829--1843, 2021.

\bibitem{interspeech12}
G.~Yu \emph{et~al.}, ``Dual-branch attention-in-attention transformer for
  single-channel speech enhancement,'' in \emph{IEEE International Conference
  on Acoustics, Speech and Signal Processing (ICASSP)}, {Singapore}, 2022, pp.
  7847--7851.

\bibitem{interspeech14}
Y.~Luo, Z.~Chen, and T.~Y. oshioka, ``Dual-path rnn: efficient long sequence
  modeling for time-domain single-channel speech separation,'' in \emph{IEEE
  International Conference on Acoustics, Speech and Signal Processing
  (ICASSP)}, {Barcelona, Spain}, 2020, pp. 46--50.

\bibitem{interspeech15}
K.~Hu, R.~Pang, T.~N. Sainath, and T.~Strohman, ``Transformer based
  deliberation for two-pass speech recognition,'' in \emph{IEEE Spoken Language
  Technology Workshop (SLT)}, {Shenzhen, China}, 2021, pp. 68--74.

\bibitem{interspeech17}
K.~Wang, B.~He, and W.-P. Zhu, ``Tstnn: Two-stage transformer based neural
  network for speech enhancement in the time domain,'' in \emph{IEEE
  International Conference on Acoustics, Speech and Signal Processing
  (ICASSP)}, 2021, pp. 7098--7102.

\bibitem{interspeech18}
Z.~Zhang, B.~He, and Z.~Zhang, ``Transmask: A compact and fast speech
  separation model based on transformer,'' in \emph{IEEE International
  Conference on Acoustics, Speech and Signal Processing (ICASSP)}, 2021, pp.
  5764--5768.

\bibitem{interspeech19}
L.~Yang, W.~Liu, and W.~Wang, ``Tfpsnet: Time-frequency domain path scanning
  network for speech separation,'' in \emph{IEEE International Conference on
  Acoustics, Speech and Signal Processing (ICASSP)}, Singapore, 2022, pp.
  6842--6846.

\bibitem{interspeech22}
G.~Huang, Z.~Liu, L.~Van Der~Maaten, and K.~Q. Weinberger, ``Densely connected
  convolutional networks,'' in \emph{IEEE Conference on Computer Vision and
  Pattern Recognition (CVPR)}, {Honolulu, HI, USA}, 2017, pp. 2261--2269.

\bibitem{interspeech20}
A.~Gulati \emph{et~al.}, ``Conformer: Convolution-augmented transformer for
  speech recognition,'' in \emph{Proc. {INTERSPEECH} 2020 --
  21\textsuperscript{st} Annual Conference of the International Speech
  Communication Association}, {Shanghai, China}, 2020, pp. 5036--5040.

\bibitem{interspeech21}
E.~Kim and H.~Seo, ``Se-conformer: Time-domain speech enhancement using
  conformer,'' in \emph{Proc. {INTERSPEECH} 2021 -- 22\textsuperscript{nd}
  Annual Conference of the International Speech Communication Association},
  {Brno, Czech}, 2021, pp. 2736--2740.

\bibitem{interspeech23}
C.~V. alentini Botinhao, X.~Wang, S.~Takaki, and J.~Yamagishi, ``Investigating
  rnn-based speech enhancement methods for noise-robust text-to-speech,'' in
  \emph{Proc. {SSW} 9\textsuperscript{th} ISCA Workshop on Speech Synthesis
  Workshop}, 2016, pp. 146--152.

\bibitem{interspeech25}
O.~Ronneberger, P.~Fischer, and T.~Brox, ``U-net: Convolutional networks for
  biomedical image segmentation,'' in \emph{International Conference on Medical
  Image Computing and Computer Assisted Intervention (MICCAI)}, {Munich,
  Germany}, 2015, pp. 234--241.

\bibitem{interspeech26}
X.~Hu \emph{et~al.}, ``Speech separation using an asynchronous fully recurrent
  convolutional neural network,'' in \emph{Advances in Neural Information
  Processing Systems (NeurIPS)}, 2021, pp. 22\,509--22\,522.

\bibitem{interspeech27}
S.~Woo, J.~Park, J.-Y. Lee, and I.~S. Kweon, ``Cbam: Convolutional block
  attention module,'' in \emph{Proceedings of the European Conference on
  Computer Vision (ECCV)}, {Munich, Germany}, 2018, pp. 3--19.

\bibitem{interspeech28}
J.~Hu \emph{et~al.}, ``Squeeze-and-excitation networks,'' \emph{IEEE
  Transactions on Pattern Analysis and Machine Intelligence}, vol.~42, no.~8,
  pp. 2011--2023, 2020.

\bibitem{interspeechsmu}
K.~Biswas, S.~Kumar, S.~Banerjee, and A.~K. Pandey, ``Smooth maximum unit:
  Smooth activation function for deep networks using smoothing maximum
  technique,'' in \emph{IEEE Conference on Computer Vision and Pattern
  Recognition (CVPR)}, {New Orleans, USA}, 2022, pp. 784--793.

\bibitem{interspeech29}
F.~Dang, H.~Chen, and P.~Zhang, ``Dpt-fsnet: Dual-path transformer based
  full-band and sub-band fusion network for speech enhancement,'' in \emph{IEEE
  International Conference on Acoustics, Speech and Signal Processing
  (ICASSP)}, Singapore, 2022, pp. 6857--6861.

\bibitem{interspeech30}
W.~Shin \emph{et~al.}, ``Multi-view attention transfer for efficient speech
  enhancement,'' in \emph{Proc. {INTERSPEECH} 2022 -- 23\textsuperscript{rd}
  Annual Conference of the International Speech Communication Association},
  {Incheon, Korea}, 2022, pp. 1198--1202.

\bibitem{interspeech36}
D.~Yin, C.~Luo, Z.~Xiong, and W.~Zeng, ``Phasen: A phase-and-harmonics-aware
  speech enhancement network,'' in \emph{Proceedings of the AAAI Conference on
  Artificial Intelligence}, vol.~34, no.~05, {New York, USA}, 2020, pp.
  9458--9465.

\bibitem{interspeech37}
J.~Lin, A.~J. d.~L. van Wijngaarden, K.-C. Wang, and M.~C. Smith, ``Speech
  enhancement using multi-stage self-attentive temporal convolutional
  networks,'' \emph{IEEE/ACM Transactions on Audio, Speech, and Language
  Processing}, vol.~29, pp. 3440--3450, 2021.

\bibitem{interspeechDemand}
J.~Thiemann, N.~Ito, and E.~Vincent, ``The diverse environments multi-channel
  acoustic noise database (demand): A database of multichannel environmental
  noise recordings,'' in \emph{Proceedings of Meetings on Acoustics}, vol.~19,
  no.~1, {Acoustical Society of America}, 2013, p. 035081.

\bibitem{interspeechvoice}
C.~Veaux, J.~Yamagishi, and S.~King, ``The voice bank corpus: Design,
  collection and data analysis of a large regional accent speech database,'' in
  \emph{2013 International Conference Oriental COCOSDA held jointly with 2013
  Conference on Asian Spoken Language Research and Evaluation
  (O-COCOSDA/CASLRE)}, 2013, pp. 1--4.

\bibitem{interspeech31}
I.~Loshchilov and F.~Hutter, ``Decoupled weight decay regularization,'' in
  \emph{International Conference on Learning Representations (ICLR)}, {New
  Orleans, USA}, 2019, pp. 1--14.

\bibitem{interspeech32}
C.~H. Taal \emph{et~al.}, ``An algorithm for intelligibility prediction of
  time-frequency weighted noisy speech,'' \emph{IEEE/ACM Transactions on Audio,
  Speech, and Language Processing}, vol.~19, no.~7, pp. 2125--2136, 2011.

\bibitem{interspeech33}
A.~Rix, J.~Beerends, M.~Hollier, and A.~Hekstra, ``Perceptual evaluation of
  speech quality (pesq)-a new method for speech quality assessment of telephone
  networks and codecs,'' in \emph{IEEE International Conference on Acoustics,
  Speech and Signal Processing (ICASSP)}, {Salt Lake City, Utah, USA}, 2001,
  pp. 749--752.

\end{thebibliography}

\end{document}